# ICSE 2022 Sustainability Report[*]

## 44th International Conference on Software Engineering, Pittsburgh, USA

Patricia Lago

Vrije Universiteit Amsterdam, p.lago@vu.nl

Marcel Böhme

Max Planck Institute for Security and Privacy (MPI-SP), Bochum, Germany, marcel.boehme@mpi-sp.org

Markus Funke

Vrije Universiteit Amsterdam, m.t.funke@vu.nl

[*]Based on the online post-conference sustainability survey.

**ABSTRACT**

The carbon footprint of academic conferences becomes a topic of increasing debate. It is important to consider whether the benefits derived from attending conferences in person outweigh the community's carbon footprint. Therefore, we need to evaluate the overall ecological consequences in relation to the perceived advantages. To that extent, we conducted a post-conference questionnaire survey among participants of the 44th International Conference on Software Engineering (ICSE) 2022 in  Pittsburgh, USA, seeking their feedback about the conference and experience from a sustainability perspective. In total, 53 participants filled out our survey. Overall, 8 of 42 respondents felt that the community's carbon footprint was not offset by the benefits of in-person attendance.

## 1  Introduction

In order to investigate whether the benefits gained by the community outweighed the overall carbon footprint generated by attending the conference, we conducted a post-conference questionnaire survey among attendees of the 44th International Conference on Software Engineering (ICSE) 2022, held in Pittsburgh, USA. The survey was administered online after the conclusion of the conference. Our objective was to gather feedback from participants on their sustainability-related experiences, which would serve as a basis for identifying areas of improvement in future editions of the conference.

The primary purpose of this report is to present the findings of the survey. Additionally, we explore potential strategies that the community can adopt to enhance the sustainability of ICSE, while maintaining the essential aspects of such academic events. Our aim is to offer guidance to future conference organizers regarding the choice of venue and the value attributed to paper presentations and social events.

## 2 Method

To design our questionnaire we used Google Forms[1] as a survey tool. The complete questionnaire encompassed 16 questions as outlined in Table 1. The feedback was collected by conducting a post-conference survey.

**Table 1: Questionnaire definition**

| ID | Question | Type |
|----|----------|------|
| 1 | *"Name"* | open text field |
| 2 | *"Affiliation"* | open text field |
| 3 | *"Country of Residence"* | open text field |
| 4 | *"Your profile"* | academic / practitioner / both / other |
| 5 | *"Attended ICSE'22"* | virtual only / in-person only / both / neither |
| 6 | *"What was (would have been) your approximate carbon footprint for traveling to Pittsburgh?"* | open text field |
| 7 | *"Considering all factors, do you feel the community's carbon footprint is offset by the community's benefits of attending ICSE in-person?"* | yes / no / maybe |
| 8 | *"From your own perspective, what are the concrete benefits of attending ICSE'22 in-person and how do they measure up against our carbon footprint?"* | open text field |
| 9 | *"ICSE'22 was organized as a hybrid conference and allowed attendees to participate both virtually and in-person. How does this hybrid model compare to the classic in-person only model?"* | open text field |
| 10 | *"Can you imagine other conference models that would maximize sustainability while maintaining many of the benefits of an in-person ICSE?"* | open text field |
| 11 | *"Imagine to be a junior participant (e.g. PhD candidate, Postdoc). From your perspective, what program activity (or activities) would be top priority for you to attend in-person and counterbalance the corresponding carbon footprint?"* | open text field |
| 12 | *"Imagine to be a senior participant (e.g. Full Professor). From your perspective, what program activity (or activities) would be top priority for you to attend in-person and counterbalance the corresponding carbon footprint?"* | open text field |
| 13 | *"In general, what program activity (or activities) would make virtual participation more attractive? And what would you certainly NOT do in a virtual program?"* | open text field |
| 14 | *"In the long term, what can we do as a community to make ICSE more sustainable?"* | open text field |
| 15 | *"In your opinion, how can we concretely measure the social benefit of an in-person (part of) ICSE to offset against our carbon footprint?"* | open text field |
| 16 | *"Is there anything more that you would like to add?"* | open text field |

---

[1] Google Forms - https://www.google.com/forms/about/

# 3    Quantitative Results

This section presents an analysis of the survey results. Despite the overall success of the 44th edition of the ICSE conference with 2025 attendees in total, our survey got a low response rate of 2.6% with 53 participants in total. The final survey data and analysis are available online[2]. As depicted in Table 1, participants were requested to provide their name and affiliation to enable us to contact them for potential follow-up inquiries. However, to ensure privacy protection, we anonymized these fields in the shared online analysis. According to the analysis in Figure 1, out of the 53 responses, 42 identified themselves as academics, 1 as practitioner, and 10 as both.

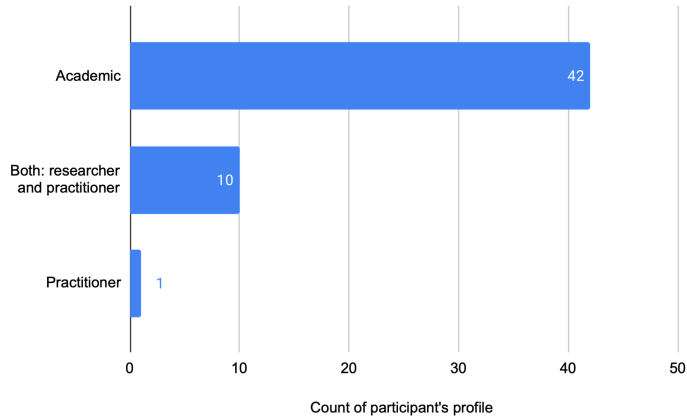

**Figure 1: Participant's profiles**

In order to highlight and raise awareness about the carbon footprint associated with participants' travel to and from the conference venue in Pittsburgh, USA, we requested attendees to utilize an online footprint calculator[3] to estimate their approximate carbon emissions (question ID 6). The findings indicate that participants, on average, generated a range of approximately 2 to 5 metric tons of $CO_2$ in relation to their attendance at the ICSE conference. The country of residence, depicted in Figure 2, indicates that most of the participants came from the USA – where the conference took place.

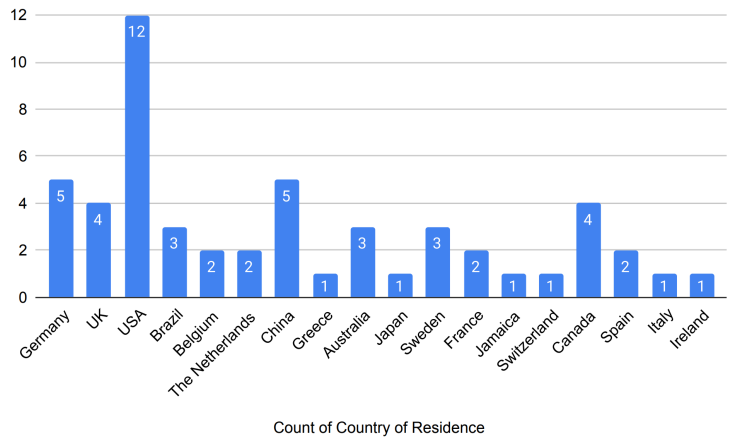

**Figure 2: Participant's country of residence**

---

Interestingly, the majority (26 participants) attended the conference only virtually as outlined in Figure 3. Only 8 attended the conference either in-person or not at all.

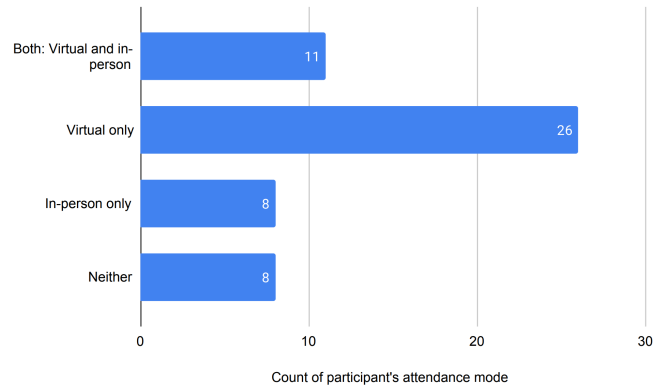

**Figure 3: Participant's attendance mode**

By asking question ID 7, we intended to perceive the participants perspective, whether the community's carbon footprint was offset by the benefits of an in-person attendance. As depicted in Figure 4, 8 out of 42 respondents answered this question with *no*, while 15 respondents felt that the footprint was offset. The remaining 29 participants are unsure.

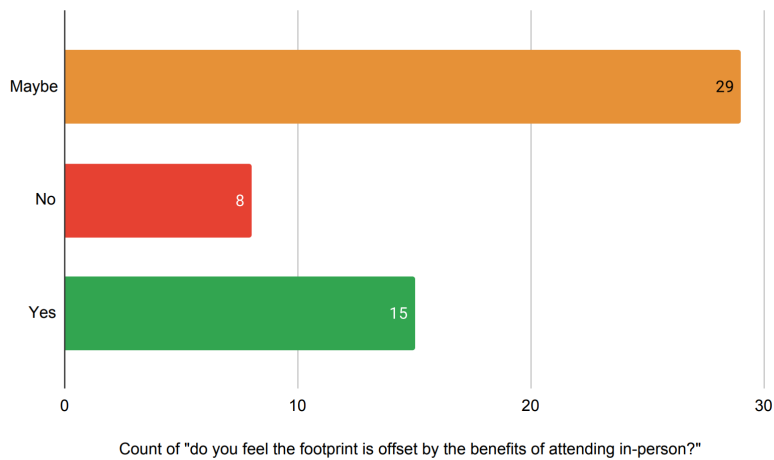

**Figure 4: Feeling of the participants whether the footprint is offset by the benefits of attending in-person**

## 4  Quantitative Results

To better understand the advantages and disadvantages of current conference modes, i.e., in-person vs. online vs. hybrid mode, we asked various open-ended questions as indicated by question ID 8 - 16. In this section, we report and summarize these qualitative insights. The complete list of answers and quotes can be found in the online survey data.

### Q-ID-8:

From your own perspective, what are the concrete benefits of attending ICSE'22 in-person and how do they measure up against our carbon footprint?

**Advantages of in-person meetings**
- *to build and grow together as a community (building online communities is not as successful), to develop a sense of belonging (relevant for newcomers/students)*
- *to energize the professional agenda of the SE community,*
- *to promote your own research vision, to talk about concrete grant ideas, paper ideas*
- *to share non-technical knowledge (e.g., supervision, teaching, etc)*
- *to provide early career individuals the opportunity to interact with experienced professionals,*
- *to promote your own PhD students (academic job market)*
- *to advertise open positions (e.g., when choosing PostDoc advisor personal meetings are always better)*
- *to stay in touch with (former) collaborators and colleagues*
- *to get to know new members of the community (students, postdocs, faculty)*
- *to observe and learn about and from \*other\* people, countries and cultures, to hear and be heard.*
- *to hear and be heard*
- *to break work/office/life routine, have fun, screen break, social events*

**Advantages of virtual meetings**
- *Better sustainability*
- *No travel costs*
- *No barriers*
- *Some people have constraints on their traveling: visa, budget, parenthood/care taking, health, limited mobility)*
- *People can attend more conferences without getting exhausted.*
- *Contribute to overall growth of our community*
- *Given the availability of teleconferencing technologies, in person events are increasingly difficult to justify (particularly against our carbon footprint)*

### Q-ID-9:

ICSE'22 was organized as a hybrid conference and allowed attendees to participate both virtually and in-person. How does this hybrid model compare to the classic in-person only model?

**The majority of respondents were in favor of a hybrid setup**
- *Best of both worlds: Wider virtual attendance from people who can normally not afford to attend, smaller physical attendance from those that want to enjoy the benefits of in-person. People can select the most relevant hybrid conference to attend in-person.*
- *Hybrid is more inclusive / improved accessibility (visa, budget, parenthood/care taking, health, limited mobility)*

- *More flexible and more friendly*

**Others raised concerns about the virtual component**
- *Unequal experience -- strictly worse experience for those who are online.*
- *Funding agencies might stop proving funding for in-person if virtual attendance is an option.*
- *The 2022 rolling time zone model was brutal and most sessions had minimal attendees.*
- *Hybrid takes weeks (almost a month in ICSE'22) of participation. We don't have that time.*
- *Limited active virtual participation. Impossible to enjoy the immersive experience of an in-person conference (incl. stumbling on acquaintances, having extra time to ask speakers after the talks). "I basically didn't do anything in the virtual space, and I ended up missing the keynotes because I hadn't realized those were virtual only."*

**Proposals (verbatim)**
- *I would go to hybrid in the same week, not in separate weeks as it was this year.*
- *Hybrid events work well if there is a clear strategy for what happens on-site and what is allocated on-line. To try to serve both groups with a similar experience is, in my experience, a bad approach (and a very difficult one).*
- *I think it would be nice for people to "virtually" present at physical meetings (e.g., via Zoom terminals that are located at the coffee/break venues).*
- *I really really think we need to ask the question as "our software engineering conferences" rather than just ICSE. There are a lot of things that even in all in person events were wasted, for example the promotion materials, call for paper prints. I also think there is not only the sustainability aspect, but also the diversity and inclusion aspect of hybrid model where we can be more inclusive of smaller organizations in a hybrid model.*

## Q-ID-10:

Can you imagine other conference models that would maximize sustainability while maintaining many of the benefits of an in-person ICSE?
- *Hub model: Regional in-person meetings at the same time (e.g., European ICSE Hub) to reduce distance traveled and maximize utility for researchers in the region. Several hubs or one single hub + everything else virtual.*
- *Co-location with other major conferences (ASE/ICSE/ISSTA) to reduce number of trips (federation model like USENIX)*
- *Co-location of smaller conferences with larger conferences to reduce the number of trips.*
- *Hybrid (or in-person first) every other year, alternating with virtual.*
- *Technical program virtual, workshops, tutorials, doctoral symposium, etc in-person to maximize community building efforts.*
- *Find the least carbon-imprinting location for everyone to travel to. Avoid places difficult to reach.*
- *Journals as main publication venues, conferences for in-person meetings.*
- *Zoom/video terminals at coffee/break venues.*

## Q-ID-11:

Imagine to be a junior participant (e.g. PhD candidate, Postdoc). From your perspective, what program activity (or activities) would be top priority for you to attend in-person and counterbalance the corresponding carbon footprint?

- *Maybe participating in a fórum to discuss academic carrier*
- *Networking for opportunities (post-docs, teaching position, researcher contracts, etc)*
- *Student mentoring, building up a network,*
- *meeting senior researchers in the community.*
- *Doctoral symposium/Early career event, social events, a workshop or collocated event dedicated to the scope of work, in addition to the conference.*
- *tutorials, workshops, networking sessions(new researchers forum)*
- *Attendance at in-person workshops, which offer the opportunities of new collaborations and friendships.*
- *For PhD candidates, the top priority is going to be advertising their work and making connections. It's going to be from people seeing their talk and then approaching them in the hall afterward, or from people chatting with them during poster session.*
- *Flagship conference(s) in my area. It's key to go to the large conferences rather than the smaller events.*

### **Q-ID-14:**

In the long term, what can we do as a community to make ICSE more sustainable?
- *Encourage participants registering for in-person to look at the CO2 costs.*
- *Be smart about choosing the next conference location (e.g., good rail options, shortest flights, etc)*
- *Producing less waste.*
  - *no plastic for lunch and drinking,*
  - *no need for useless gadgets,*
  - *re-cycling conference badges,*
  - *choose responsible catering service and venues.*
  - *going paperless (no reception etc tickets on paper);*
  - *no conference flyers or advertisment brochures. Instead, provide advertisement opportunity on the website, emails, chats, etc),*

## 5  Conclusion

Through this post-conference online questionnaire survey, our aim was to assess whether the benefits of in-person attendance at the 44th ICSE edition outweighed the carbon emissions generated. However, the study's limited response rate constrains our ability to draw definitive conclusions or perform statistical analysis. Nonetheless, the inclusion of open-ended questions in our survey provided valuable insights into participants' perspectives on academic conferences and potential future conference models.

The majority of respondents mention social factors as advantages of in-person meetings, such as networking, advertising their research, and providing feedback for early career researchers. We believe that future conference models need to take the social aspects more seriously by setting the focus on networking aspects. It can also be concluded that attending conferences in-person for early career researchers, such as PhD candidates can have a major impact on their career development. Based on the suggestions provided by our survey participants, we recommend incorporating more social opportunities for them, such as networking sessions, forums for new researchers, workshops, and tutorials.

In the future, conference organizers should explore alternative conference models that go beyond the traditional in-person format. As highlighted by our survey participants, the hub model offers benefits on multiple fronts: reducing the carbon footprint associated with long-distance travel by minimizing the number of intercontinental flights, while still facilitating valuable networking and social interactions through in-person meetings for early career researchers. By adopting such an approach, conferences can achieve a balance between sustainability and preserving the important social aspects of academic gatherings.

As a future direction, it is recommended that forthcoming editions of the ICSA conference incorporate an assessment of the correlation between environmental considerations and perceived value of the conference. This can be accomplished by conducting on-site surveys to gather feedback regarding various session types. This evaluation would enable a conclusion to be drawn regarding whether the environmental impacts caused by transportation and other factors are justified by the value attributed to traditional research paper presentations.